\documentclass[11pt]{article} % use larger type; default would be 10pt

\usepackage{fullpage}
\usepackage[utf8]{inputenc} % set input encoding (not needed with XeLaTeX)
\usepackage{amsmath,amsthm,amsfonts}    
\usepackage{color}
\usepackage[usenames,dvipsnames]{xcolor}
\usepackage{url}
\usepackage[colorlinks=true,breaklinks, linkcolor=Gray, citecolor=Gray, urlcolor=Gray]{hyperref}

\newcommand{\ket}[1]{\left\vert{#1}\right\rangle}

\newcommand{\eqnref}[1]{\hyperref[#1]{{(\ref*{#1})}}}
\newcommand{\thmref}[1]{\hyperref[#1]{{Theorem~\ref*{#1}}}}
\newcommand{\lemref}[1]{\hyperref[#1]{{Lemma~\ref*{#1}}}}
\newcommand{\corref}[1]{\hyperref[#1]{{Corollary~\ref*{#1}}}}
\newcommand{\defref}[1]{\hyperref[#1]{{Definition~\ref*{#1}}}}
\newcommand{\secref}[1]{\hyperref[#1]{{Section~\ref*{#1}}}}
\newcommand{\figref}[1]{\hyperref[#1]{{Figure~\ref*{#1}}}}
\newcommand{\tabref}[1]{\hyperref[#1]{{Table~\ref*{#1}}}}
\newcommand{\remref}[1]{\hyperref[#1]{{Remark~\ref*{#1}}}}
\newcommand{\appref}[1]{\hyperref[#1]{{Appendix~\ref*{#1}}}}
\newcommand{\claimref}[1]{\hyperref[#1]{{Claim~\ref*{#1}}}}
\newcommand{\propref}[1]{\hyperref[#1]{{Proposition~\ref*{#1}}}}
\newcommand{\exampleref}[1]{\hyperref[#1]{{Example~\ref*{#1}}}}
\newcommand{\conjref}[1]{\hyperref[#1]{{Conjecture~\ref*{#1}}}}

\usepackage{graphicx}
\usepackage{subcaption}

\title{Quantum circuit optimization by topological compaction in the surface code}
\author{Adam Paetznick\\ David R. Cheriton School of Computer Science, and \\ Institute for Quantum Computing,\\ University of Waterloo
\and
Austin G. Fowler\\
Centre for Quantum Computation and Communication Technology,\\
School of Physics,\\ 
The University of Melbourne
}
\date{} % Activate to display a given date or no date (if empty),
\begin{document}
\maketitle

\begin{abstract}
The fragile nature of quantum information limits our ability to construct large quantities of quantum bits suitable for quantum computing. An important goal, therefore, is to minimize the amount of resources required to implement quantum algorithms, many of which are serial in nature and leave large numbers of qubits idle much of the time unless compression techniques are used.  Furthermore, quantum error-correcting codes, which are required to reduce the effects of noise, introduce additional resource overhead.

We consider a strategy for quantum circuit optimization based on topological deformation in the surface code, one of the best performing and most practical quantum error-correcting codes.  Specifically, we examine the problem of minimizing computation time on a two-dimensional qubit lattice of arbitrary, but fixed dimension, and propose two algorithms for doing so. 
\end{abstract}

\section{Introduction}
The task of building a large-scale quantum computer is a challenging one. Quantum information tends to decohere quickly, and thus our ability to construct and manipulate quantum bits (qubits) in large quantities is limited. An important goal, therefore, is to minimize the number of qubits and the amount of time required to implement quantum algorithms.  However, many quantum algorithms are serial in nature, leaving large numbers of qubits idle much of the time.  
Low-gate-count arithmetic quantum circuits, for example, form a staircase structure of linear depth~\cite{Cuccaro2004}.
Parallelization of certain procedures, such as the quantum Fourier transform, is possible when extra qubits are available but is typically done on a case by case basis~\cite{Cleve2000a}.

Error-correcting codes are required in order to reduce the impact of decoherence and faulty control of the computer~\cite{Shor1997}.  While it is possible introduce error-correction without inducing time overhead~\cite{Fowler2012g}, any fault-tolerant circuit will require an increase in either time or space as compared to the corresponding ideal quantum circuit. The resource increase can be very large, often as much as a million fold or more~\cite{Knill2004,Raussendorf2007a,Paetznick2011,Jones2012b}.  Resource estimates are even larger if geometric constraints of the quantum computer are considered.
Indeed, many proposals for quantum computers impose a fixed-size two-dimensional lattice, with limited interactions between qubits~\cite{Levy2001,Weinstein2005,Devi08,Helmer2009,Divincenzo2009,Amin10,Jones2012b,Kump11,Levy2011,Kloeffel2013}.

We propose an automated and global resource optimization solution which is fault-tolerant and accounts for geometry and locality constraints by operating within the surface code.  
Our strategy is to minimize computation time by smoothly reshaping the computation in order to eliminate wasted space.
Two algorithms are presented. The first is a force-directed algorithm in which the fault-tolerant quantum circuit is treated as a malleable physical object. The second algorithm is based on simulated annealing.
Each algorithm takes as input an ideal quantum circuit and a fixed-size two-dimensional qubit lattice and outputs a corresponding compact fault-tolerant quantum circuit that operates entirely within the lattice dimensions.

A variety of techniques have been proposed for reducing resource overhead in fault-tolerant quantum circuits.  For schemes based on concatenated codes, the overhead is dominated by error correction circuits which require carefully prepared encoded states.  Efficient techniques for preparing such states have been developed and continue to improve~\cite{Steane2004,Reichardt2006,Paetznick2011}.  Another significant source of overhead is the execution of particular gates that are required for universality.  These gates involve preparation and distillation of exotic resource states~\cite{Bravyi2004}.  Recently there have been a number of proposals for improving the efficiency of distillation~\cite{Meier2012,Bravyi2012a,Jones2012c,Fowler2012f,Fowler2013}.

The above techniques have been effective at reducing the resources required by fault-tolerant quantum circuits.  However, they are largely manual, and focus on a small but repeated part of the circuit. 
They do not address, for example, global parallelism concerns.

Some techniques for pararallelization exist. Typical methods involve local circuit rewriting rules for trading between sequences of gates and additional qubits~\cite{Moore1998,Maslov2008,Saeedi2010}.
Small-depth circuits can be achieved for certain sub-classes of quantum circuits.  Clifford group circuits, for example, can be parallelized to quantum circuits of constant depth followed by log-depth classical post-processing~\cite{Raussendorf2001}.

Others have proposed global circuit optimization procedures that involve a multi-staged transformation to and from the measurement-based quantum computing model~\cite{Broadbent2007,DaSilva2013}. Indeed, there are strong similarities between the measurement-based model and the surface code~\cite{Raussendorf2007}.
However, the template-based and measurement-based optimizations are not fault-tolerant and, except for~\cite{Saeedi2010}, do not explicitly consider geometric constraints imposed by the quantum computer.  It is not clear that the resulting circuits remain compact under such restrictions.

Alternatively, since our algorithms operate within the surface code, the output is automatically fault-tolerant and can be easily mapped to a wide variety of two-dimensional nearest-neighbor architectures~\cite{Divincenzo2009,Ghosh2012}.
Furthermore, the rules for topologically transforming surface code braids are conceptually  simple.  There is no need to break up the transformation into multiple stages.
Thus, compared to other proposals, we feel that our approach is easier to understand, implement, and extend.

That is not to say that braid compaction is \emph{computationally} easy, however.
In fact, based on known complexities of other similar problems such as VLSI placement~\cite{Schlag1983} and container loading (see, e.g.,~\cite{Scheithauer}), we conjecture that the problem of minimizing the time (height) of a braid on a fixed rectangular lattice of qubits is NP-complete.  Our optimization algorithms are therefore crafted from carefully designed heuristics in order to handle large-scale instances. 

The first algorithm is loosely based on the physical principles of gravity and tension. Roughly, the braid is treated as a heavy tangle of rubber bands that is allowed to slide into a box under the force of gravity.  The idea is vaguely analogous to techniques used in graph drawing, for example~\cite{Kobourov2012}. The second algorithm uses the technique of simulated annealing.  In simulated annealing the solution landscape is explored by making random incremental changes and favoring those changes which decrease the solution size.  Our algorithm is inspired by a similar procedure for VLSI placement~\cite{Hsieh}.

We begin with a brief summary of the surface code in~\secref{sec:surface-code} followed by a more formal statement of the braid compaction problem in~\secref{sec:braid-compaction}.  In~\secref{sec:iterative-forcing} we describe the force-directed algorithm in detail, and a C++ implementation that we call Braidpack. The simulated annealing algorithm is presented in~\secref{sec:simulated-annealing}.

\section{The surface code}
\label{sec:surface-code}

In this section, we give a brief pedagogical introduction to the surface code.  The focus here is on the mapping from a quantum circuit to a surface code braid.  Other details of the surface code are not essential for understanding our braid compaction algorithms.
For a comprehensive introduction to the surface code we refer the reader to~\cite{Fowler2012d}.

The surface code has a number of desirable properties.
First, it operates on a two-dimensional rectangular lattice of qubits.  All operations can be performed using only one-qubit gates, and two-qubit gates involving only nearest neighbor qubits.  As a result, the required number of qubits scales much more slowly for the surface code than for concatenated codes on $2$-D nearest-neighbor architectures.  At the same time, the surface code tolerates noisier physical gates than many other quantum error correcting codes.  Reliable computation is possible so long as the noise rate is below roughly $0.6$ percent per gate.

Error-correction is performed by periodically measuring four-qubit operators, known as stabilizers, around each face and and each vertex of the lattice.  Codewords correspond to the simultaneous $\pm 1$-eigenstates of the operators.  Each measurement imposes a restriction on the Hilbert space of the lattice. Encoded qubits are created by disabling some of the measurements, thereby adding new degrees of freedom. 
We choose to define a qubit as a pair of \emph{defects}.  Defects are contiguous regions of the lattice for which the stabilizers are \emph{not} measured. 
There are two types of defects, primal and dual.  Primal defects correspond to operators around vertices of the lattice, and dual defects correspond to operators around the faces of the lattice.

Error protection is achieved by creating defects of sufficient size, and by keeping defects well separated in space.  For a code distance of $d$, we require that all defects have circumference $d$ and that defects of the same type are separated in $L_\infty$ distance by $d$. For defects of opposite type, the minimum distance depends on the shape of each defect. In all cases a distance of $d/4$ is sufficient, though in some cases primal and dual defects may be as close as $d/8$.

\begin{figure}
\centering
\begin{subfigure}[b]{.45\linewidth}
	\centering
	\includegraphics[width=\linewidth]{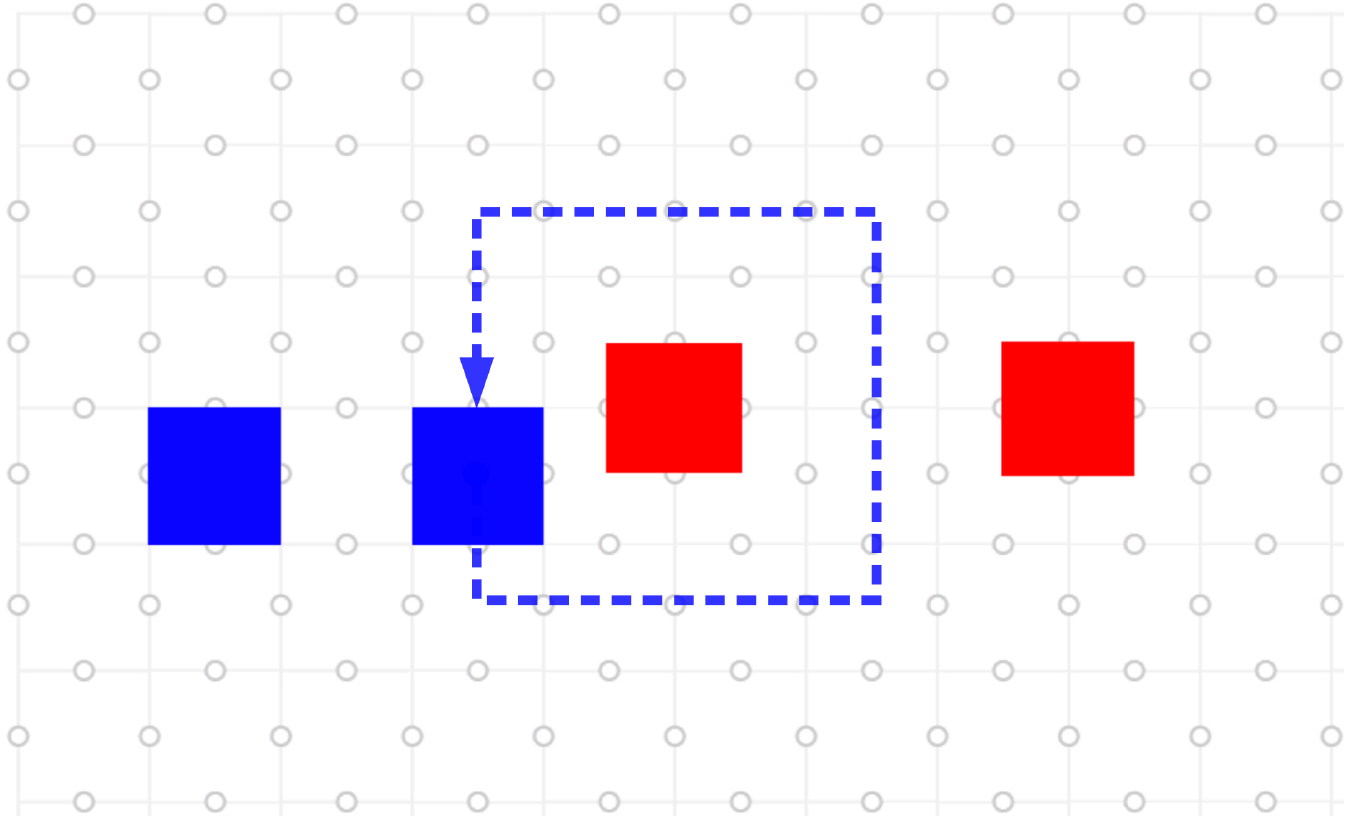}
	\caption{top view}
\end{subfigure}
\begin{subfigure}[b]{.45\linewidth}
	\centering
	\includegraphics[width=.7\linewidth]{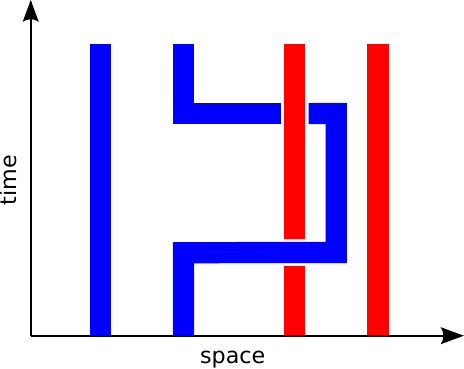}
	\caption{side view}
\end{subfigure}
\caption{\label{fig:primal-dual-cnot}
(a) Encoded surface code qubits are defined by pairs of defects, either primal (red) or dual (blue). Each defect is composed of multiple physical qubits on the two-dimensional lattice. Operations are performed by moving defects around.  Here, an encoded two-qubit operation is performed by moving one defect from the dual encoded qubit around one of the defects of the primal encoded qubit. (b) The same operation can be written as a space-time diagram in which one of the space axes has been flattened.
}
\end{figure}

Most encoded operations in the surface code proceed by moving defects around each other.  
Defect movement is achieved by turning off new regions of stabilizer measurements and then turning on other stabilizer measurements.
The movement can be divided into time-slices.
By stacking time-slices on top of each other, the encoded operations are represented by a three-dimensional object in space and time called a \emph{braid}. See~\figref{fig:primal-dual-cnot}.
Transformation of a quantum circuit to a braid can be done systematically by constructing canonical braid elements for each quantum gate.  Preparation of encoded $\ket 0$  is represented by a ``U''-shaped primal defect.  Encoded $Z$-basis measurement is essentially the reverse.  A controlled-NOT (CNOT) operation is performed by a loop of dual defects that wraps around the two associated encoded qubits. See~\tabref{tab:canonical-gate-set}.

\begin{table}
\centering
\def\imagetop#1{\vtop{\null\hbox{#1}}}
\begin{tabular}{c@{\qquad}c@{\qquad}c@{\qquad}c@{\qquad}c}
\includegraphics[width=.05\linewidth]{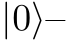}&
\includegraphics[width=.05\linewidth]{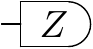}&
\includegraphics[width=.05\linewidth]{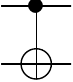}&
\includegraphics[width=.05\linewidth]{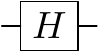}&
\includegraphics[width=.05\linewidth]{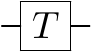} 
\\
\imagetop{\includegraphics[width=.04\linewidth]{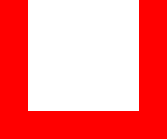}}&
\imagetop{\includegraphics[width=.04\linewidth]{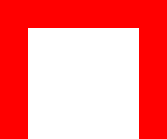}}&
\imagetop{\includegraphics[width=.08\linewidth]{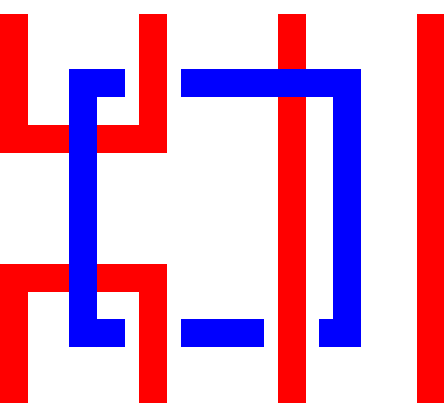}}&
\imagetop{\includegraphics[width=.04\linewidth]{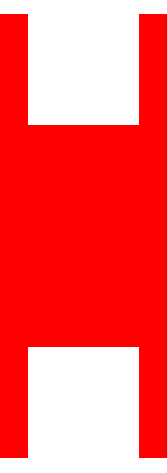}}&
\imagetop{\includegraphics[width=.2\linewidth]{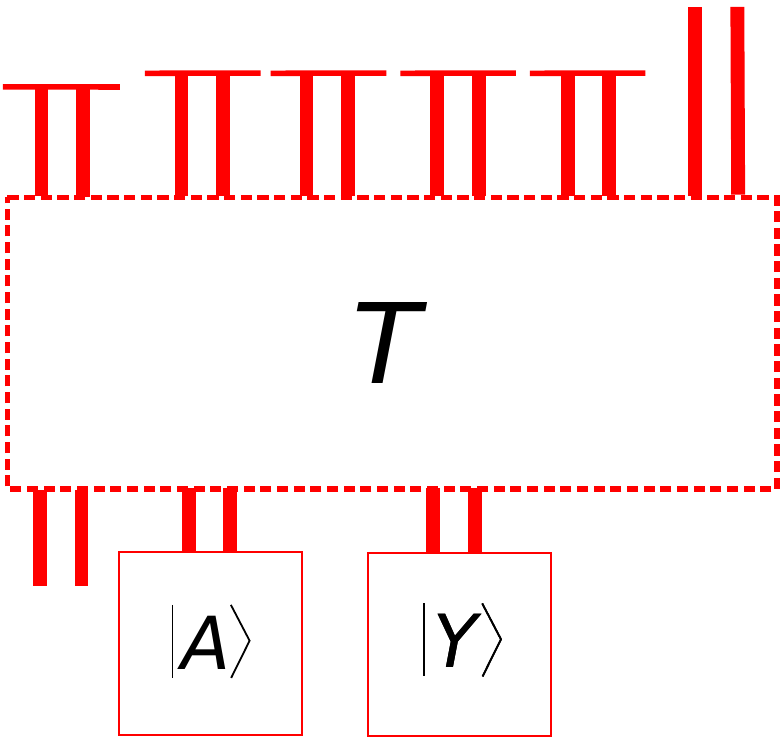}}
\end{tabular}
\caption{\label{tab:canonical-gate-set}
The surface code gate set (top) and corresponding canonical braids (bottom).  Each braid is a three-dimensional collection of defects.  For visual clarity, the braids have been flattened here into two dimensions.  
}
\end{table}

\begin{figure}
\centering
\begin{subfigure}[b]{.15\linewidth}
	\includegraphics[width=.8\linewidth]{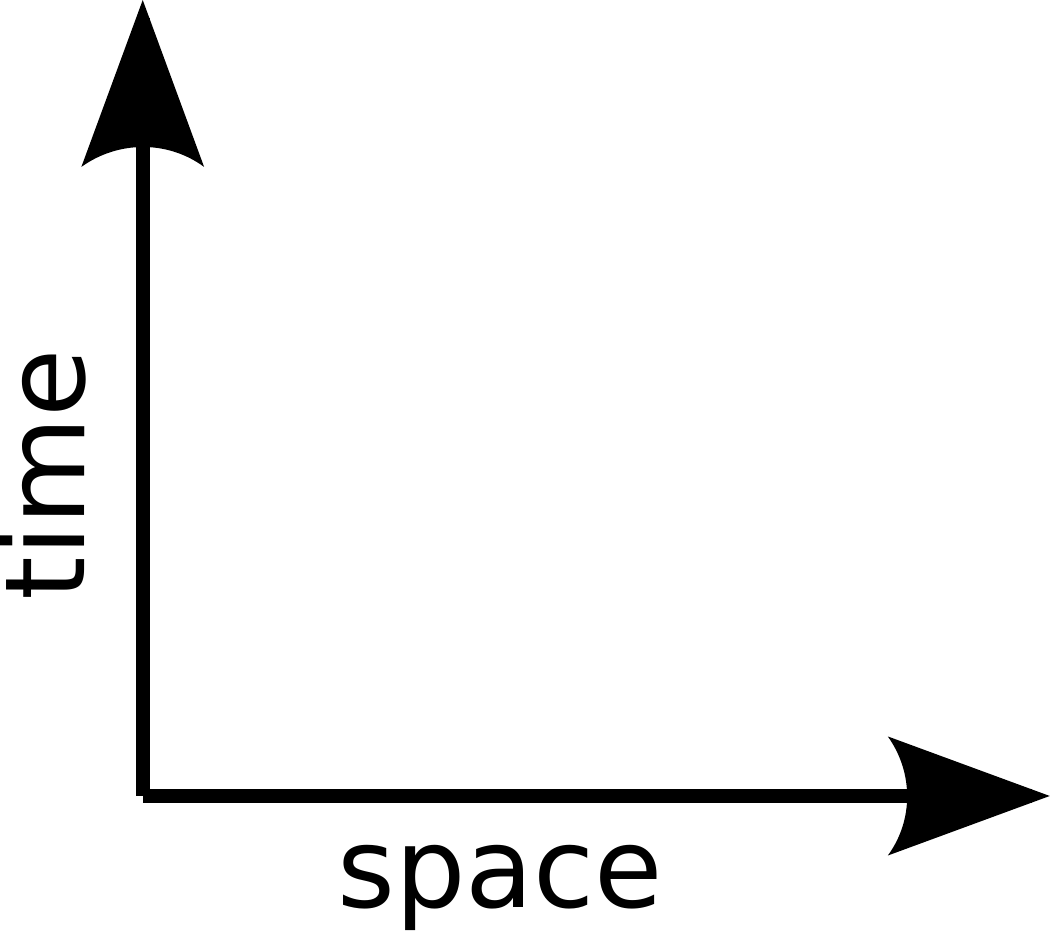}
\end{subfigure}
\begin{subfigure}[b]{.25\linewidth}
	\centering
	\includegraphics[height=2.5cm]{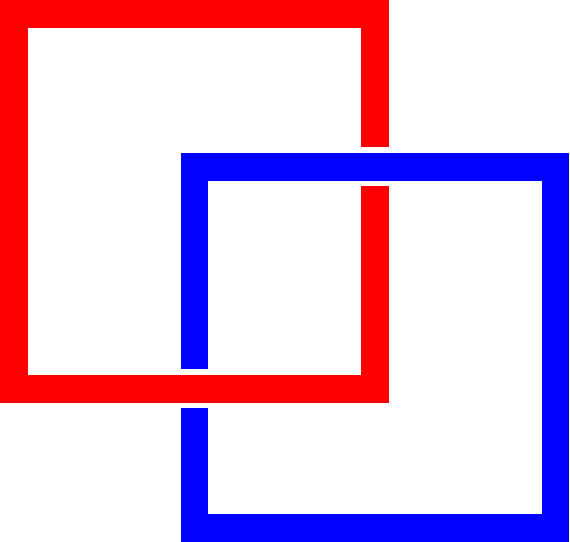}
	\caption{}
\end{subfigure}
\hspace{.1\linewidth}
\begin{subfigure}[b]{.25\linewidth}
	\centering
	\includegraphics[height=2.5cm]{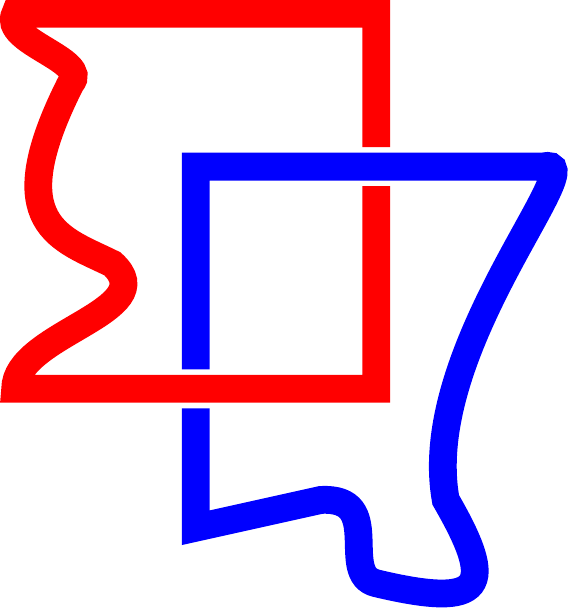}
	\caption{}
\end{subfigure}
\caption{\label{fig:topological-deformation} Surface code braids are invariant under topological deformation.  The space-time diagram on the left (a) is topologically equivalent to the diagram on the right (b). Defect strings and loops may be smoothly stretched and contracted without altering the encoded operation.
}
\end{figure}

Braids consisting of these operations are invariant under topological deformation.  That is, a quantum circuit can be represented by a canonical braid, and also by any braid that is topologically equivalent to that canonical braid.  Strings of defects may be smoothly pulled or pushed around in space and time without altering the encoded quantum computation. See~\figref{fig:topological-deformation}.  Note that space and time are symmetric here.  Space can be traded for time and vice versa.

Not all encoded operations in the surface code can be performed topologically, however.  The encoded Hadamard operation, for example, requires the encoded qubit---i.e., the two corresponding defects---to be placed on a separate lattice, isolated from all other encoded qubits.  This is achieved by first ``cutting out'' part of the lattice around the encoded qubit and then later re-attaching it to the rest of the lattice~\cite{Fowler2012b}. The resulting space-time volume is a cuboid (i.e., a box) of dimension roughly $3d/2 \times 3d/2 \times 5d/2$.  
However, the cuboid contains a variety of boundary types near the surface, thus imposing some restrictions on the configurations of other surrounding defects.
The cuboid can be translated in any direction, or rotated about the time-axis by increments of $\pi/2$, but is otherwise treated as a rigid object.\footnote{In principle, a sideways Hadamard gate is possible and would allow for rotations about the $x$ and $y$ axes.  However, the chosen implementation requires the cuboid to be vertically oriented.}
For concreteness, we adopt the convention that time corresponds to the $z$-axis.

There is one other non-topological operation, the encoded $T$-gate which performs the rotation
$\left( \begin{smallmatrix}
1 & 0 \\ 
0 & e^{i\pi/4}
\end{smallmatrix} \right)$.  
This gate cannot be implemented directly in the surface code and is instead constructed by a process known as state injection and distillation followed by gate teleportation~\cite{Bravyi2004}.  It does not explicitly require the encoded qubit to be cut out of the lattice, as the Hadamard does.  However, both the distillation and gate teleportation involve measurements which are probabilistic. The required circuit changes depending on the measurement outcomes. See~\figref{fig:tgate-injection-circuit}.

Likewise, the corresponding braid cannot be entirely determined ahead of time.
It is possible, however, to shift all of the non-determinism either offline or into logical measurements, which can be performed very efficiently~\cite{Fowler2012g}.  \figref{fig:tgate-time-optimal} shows an alternative circuit that also implements $T$.  In this circuit, an $S = \left(\begin{smallmatrix}1&0\\0&i\end{smallmatrix}\right)$ gate, implemented with the help of a resource state $\ket Y = \frac{1}{\sqrt 2}(\ket 0 + i\ket 1)$, is selectively teleported into the circuit conditioned on the outcome of an $Z$-basis measurement.  Given states $\ket A$ and $\ket Y$, the entire circuit is determined ahead of time except for the measurement bases for selective teleportation.

The circuit in~\figref{fig:tgate-injection-circuit} is smaller than that of~\figref{fig:tgate-time-optimal}.  The latter circuit, however, has the advantage that it can be composed in parallel with any number of additional $T$ gate circuits.  The braid corresponding to the single-qubit unitary $THT$, for example, can be parallelized as shown in~\figref{fig:tgate-time-ordering}. The logical measurements in this braid are implemented differently than previously discussed.  The cap on the defects has been flattened into a wider, but thinner set of defects that looks like a tabletop.  This allows for maximum parallelization of sequences of $T$ gates.

The measurement regions of~\figref{fig:tgate-time-ordering} must obey a relative time ordering.
In particular, the $Z$-basis measurement of the input qubit $\ket\psi$ must be completed before the selective teleportation measurements can be performed.  In addition, the selective teleportation of the \emph{previous} $T$ (if applicable) must be completed before selective teleportation measurements of current $T$ gate can be performed.
In this way, the measurement regions for sequences of $T$ gates form a tree.  Each measurement region must be located strictly later in time than each of its children.

There are a variety of options for preparing the $\ket A$ and $\ket Y$ states required by~\figref{fig:tgate-time-optimal}.  The $\ket A$ state, for example, can be prepared by a $15$:$1$ injection and distillation procedure due to Bravyi and Kitaev~\cite{Bravyi2004}, or one
of several recent proposals~\cite{Meier2012,Bravyi2012a,Jones2012c}.
Efficient surface code braids are known for several of these protocols~\cite{Fowler2012f,Fowler2013}, though we will not discuss the details here.  Rather, for simplicity we abstract the $\ket A$ and $\ket Y$ preparation as rigid cuboids, similar to the Hadamard gate.  This gives us the freedom to define braid compaction algorithms without being coupled to a particular distillation procedure.

\newsavebox{\Smatrix}
\savebox{\Smatrix}{$\left(\begin{smallmatrix}1&0\\0&i\end{smallmatrix}\right)$}

\begin{figure}
\centering
\begin{subfigure}[b]{.4\linewidth}
  \centering
  \includegraphics[width=\linewidth]{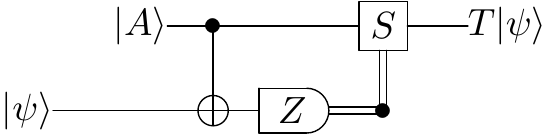}
  \caption{\label{fig:tgate-injection-circuit}}
\end{subfigure}
\hfill
\begin{subfigure}[b]{.5\linewidth}
  \centering
  \includegraphics[width=\linewidth]{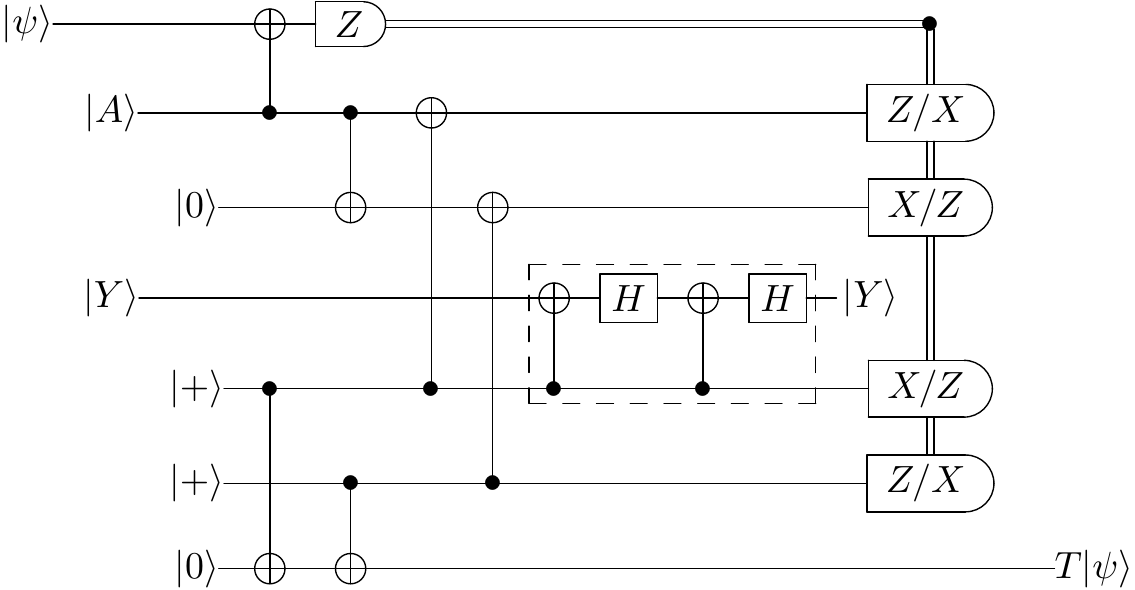}
  \caption{\label{fig:tgate-time-optimal}}
\end{subfigure}
\caption{
Two circuits that implement the $T$ gate on input state $\ket\psi$. (a) The resource state $\ket A = \ket 0 + e^{i \pi/4}\ket 1$ is constructed by injection and distillation.  Conditioned on the measurement outcome, a corrective 
$S = \usebox{\Smatrix}$ rotation may be required, which requires a non-destructive use of an ancilla $\ket Y = \ket 0 + i\ket 1$ state, initially prepared by injection and distillation (not shown).
(b) Instead of performing the conditional $S$ gate directly, selective destination teleportation can be used~\cite{Fowler2012g}.  On one path of the teleportation, the $S$ gate is applied, and on the other path it is not.  The $Z$-basis measurement on $\ket\psi$ determines the bases in which the other four qubits are measured.  The output is $T\ket\psi$, up to Pauli corrections from teleportation.}
\end{figure}

\begin{figure}
\centering
\begin{subfigure}[b]{.25\linewidth}
	\centering
	\includegraphics[width=.9\linewidth]{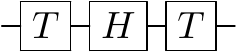}
	\caption{}
\end{subfigure}
\begin{subfigure}[b]{.99\linewidth}
	\centering
	\includegraphics[width=.75\linewidth]{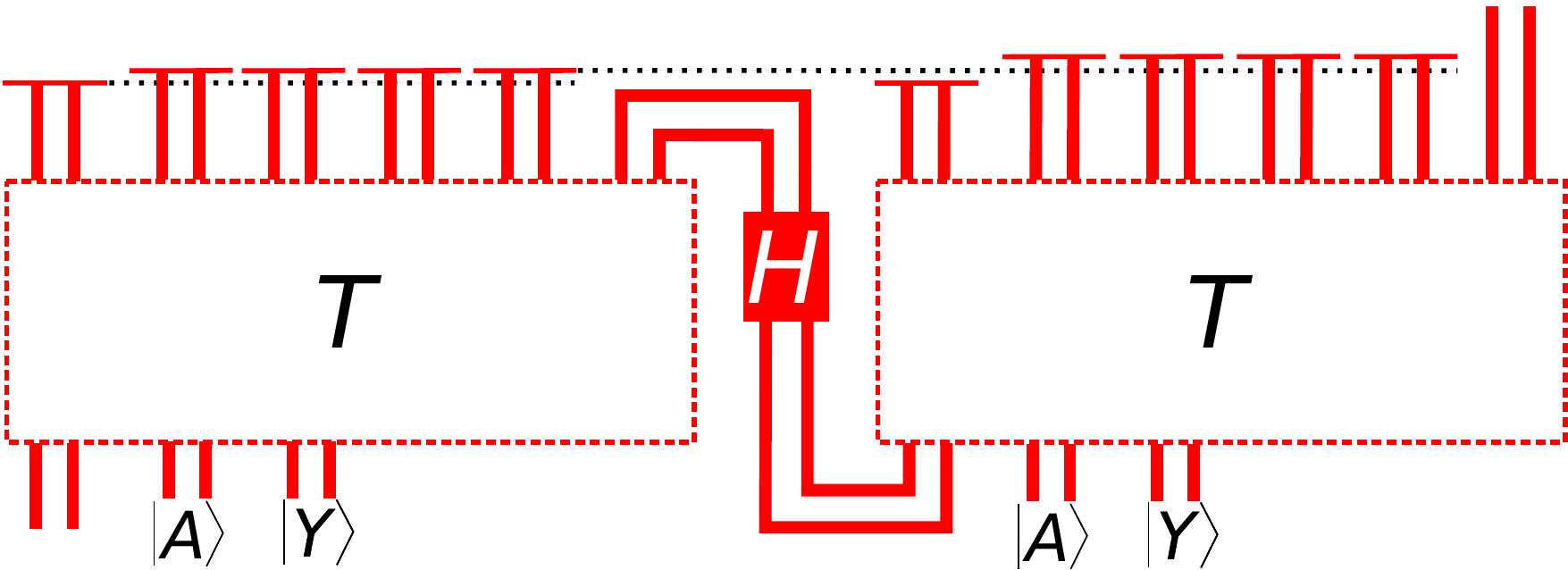}
	\caption{}
\end{subfigure}
\caption{\label{fig:tgate-time-ordering}
(a) A quantum circuit for the single-qubit unitary $THT$ in which time runs left to right. (b) A schematic representation of the corresponding surface code braid in which time runs bottom to top.  For simplicity, the braids corresponding to~\figref{fig:tgate-time-optimal} are shown as boxes, except for the measurements which are shown as thin tabletop structures.  The $T$ and $H$ boxes may be placed in parallel, and the $\ket A$ and $\ket Y$ states may be prepared ahead of time. The first measurement of the $T$ gate must complete before the remaining four selective teleportation measurements can be performed. Selective teleportation measurments between $T$ gates also obey a relative time-ordering as indicated by the black dotted lines.  Any sequence of single-qubit gates from $\{T,H,S\}$ may be parallelized in this way.
}
\end{figure}

The gates listed in~\tabref{tab:canonical-gate-set} are universal for quantum computing.  Thus any quantum circuit can be mapped to a surface code braid by first decomposing it into this gate set, and then sequentially constructing each of the canonical braid elements.

\section{The braid compaction problem}
\label{sec:braid-compaction}
The canonical braid is a fault-tolerant representation of the original circuit, but there is no guarantee that it will fit onto the two-dimensional lattice of qubits that is available.  Indeed, the structure of the canonical braid closely resembles that of the original circuit.  It is essentially a long line of defects that extends out in time.  Even if the braid fits, its two-dimensional shape means that most of the qubits in the quantum computer will be left unused.

Of course, one could try to compile the braid in a different way, so as to use more of the available space.  However, the efficiency of the compilation will depend heavily on the structure of the original circuit.  Qubits that were originally local when arranged linearly might be placed far apart when arranged in two-dimensions, thereby increasing the volume required for a CNOT between the two.

We instead choose to optimize the canonical braid by smoothly deforming it.   
So long as the deformations are topological, the optimized braid will be logically equivalent to the original.
Braid compaction, then, is the problem of taking a braid $B$ and converting it into a topologically equivalent braid $B'$ that fits into a smaller bounding volume. 
Alternatively, the problem can be described as follows.
\begin{description}
\item[Braid compaction] Given a braid $B$ and a rectangular lattice of dimension $A = (x,y)$, find a braid $B'$ that is topologically equivalent to $B$ and such that $B'$ is contained in a volume $V = (x,y,z)$ of minimum size.
\end{description}

The $x$ and $y$ dimensions of the bounding volume are are fixed by the size and geometry of the quantum computer. The goal is to efficiently use the provided space in order to minimize computation time.

Abstractly, we can view braid compaction as a process of placing cuboids (Hadamard and state distillation) in a large box, subject to certain distance, connectivity and topology constraints.  When viewed in this way, the problem looks strikingly similar to that of VLSI placement~\cite{Schlag1983}.  In the VLSI placement problem, the task is to pack a set of circuit elements---represented by rectangles---on a two-dimensional circuit board of minimum area.  Some of the circuit elements must be connected by wires, and some must be separated from other circuit elements by a minimum distance.

VLSI placement is NP-complete~\cite{Schlag1983}.  Given the close similarities with VLSI placement and with other packing problems, we conjecture that braid compaction is also NP-complete. However, despite their similarities, there are several key differences between VLSI placement and braid compaction.  In particular, the rigid objects in VLSI placement have arbitrary dimension whereas the Hadamard cuboids in the braid are of fixed size.  Thus a naive reduction from VLSI placement to braid compaction is not possible.  Attempts at a more complicated reduction or reduction from related problems such as $3$-Partition and bin packing have so far failed.

\section{A force-directed compaction algorithm}
\label{sec:iterative-forcing}

We now describe our force-directed algorithm, the first of two proposed algorithms for braid compaction. The algorithm employs two complementary ``forces''.  A gravity force acts to pull the braid down toward the bottom of the space-time grid, thereby reducing computation time.  Meanwhile, a tension force prevents the braid from becoming too large and impeding the progress of gravity.

\subsection{Braid representation}
\label{sec:forcing-braid-model}
\begin{figure}
\centering
\begin{subfigure}[b]{.24\linewidth}
	\centering
	\includegraphics[width=.6\linewidth]{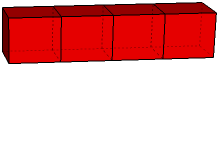}
	\caption{straight primal}
\end{subfigure}
\hfill
\begin{subfigure}[b]{.24\linewidth}
	\centering
	\includegraphics[width=.6\linewidth]{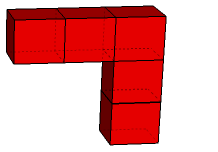}
	\caption{bent primal}
\end{subfigure}
\hfill
\begin{subfigure}[b]{.24\linewidth}
	\centering
	\includegraphics[width=.6\linewidth]{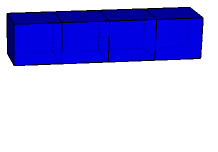}
	\caption{straight dual}
\end{subfigure}
\hfill
\begin{subfigure}[b]{.24\linewidth}
	\centering
	\includegraphics[width=.6\linewidth]{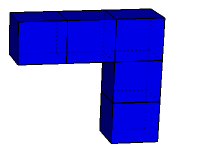}
	\caption{bent dual}
\end{subfigure}
\caption{\label{fig:pipes}
In the force-directed algorithm, braids are constructed by rotating and connecting the four primative ``plumbing'' pieces shown above.}
\end{figure}

For our force-directed algorithm, the braid is modeled as a set of plumbing pieces (i.e., pipes) placed on a three-dimensional grid.  For circuits containing preparation, measurement, single-qubit Paulis and CNOT gates, only four types of pipes are required: straight and bent (elbow shaped) pipes, both primal and dual.  See~\figref{fig:pipes}. The braid is then constructed by connecting pipes into interlocking loops.  Junctions can also be supported by merging two or more pipes.

\begin{figure}
\centering
\begin{subfigure}[b]{.1\linewidth}
  \includegraphics[scale=.4]{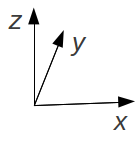}
\end{subfigure}
\begin{subfigure}[b]{.3\linewidth}
  \includegraphics[scale=.2]{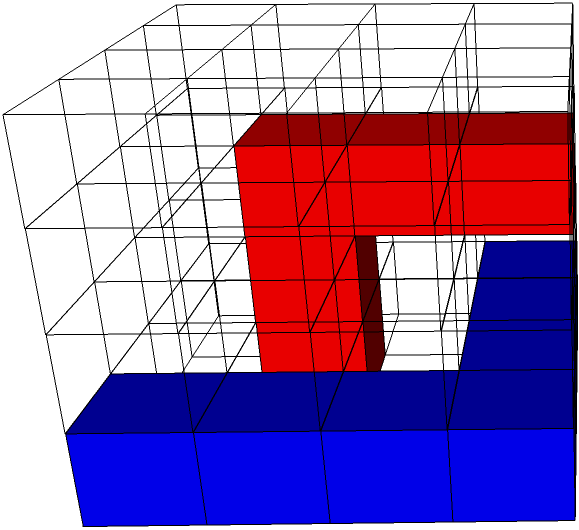}
  \caption{}
\end{subfigure}
\begin{subfigure}[b]{.3\linewidth}
  \includegraphics[scale=.2]{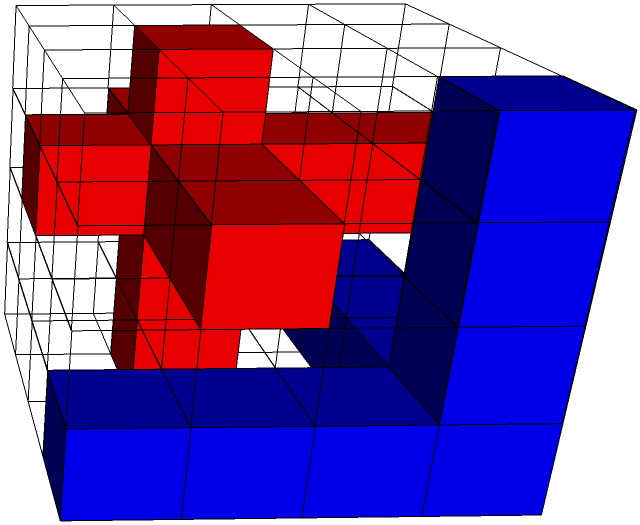}
  \caption{}
\end{subfigure}
\caption{\label{fig:cells}
(a) An example of a $4$ x $4$ x $4$ cell containing both a primal pipe and a dual pipe. The primal pipe connects to the southern face ($-z$) and the eastern face ($+x$).  The dual pipe connects to the the western face ($-x$) and to the far face ($+y$). (b) All possible pipes superimposed on a single cell.  Primal and dual defects are always separated by at least one unit cube.  Neighboring unconnected defects of the same type are always separated by at least three unit cubes.
}
\end{figure}

The three-dimensional ($l\times w \times h$) grid is partitioned into $4 \times 4 \times 4$ cells, each of which contain at most one primal pipe and one dual pipe.  Each pipe connects to at least two of the faces of the cell. For each face there is a designated unit cube to which a pipe can connect.  For example, a primal pipe that connects to the $-y$ face must always connect at position $(1,0,2)$ within the cell. Including the empty pipe, there are $2^6=64$ possible primal pipes and $64$ possible dual pipes, for a total of $4096$ possible cell configurations. See~\figref{fig:cells}.

The structure of the cell enforces a minimum distance of a single unit cube between defects of opposite type and a  distance of three unit cubes between distinct defects of the same type.  Thus, if the length of a unit cube is $\delta$, the resulting surface code distance is $d=3\delta$.  A unit cube contains $2\delta$ physical qubits per side (including qubits for stabilizer measurement), so that a single time-slice of a cell contains $64 \delta^2$ qubits. 

Regions such as Hadamards, state distillation and tabletop measurement for $T$ gates cannot be represented as a collection of conventional plumbing pieces.
Instead, they are represented by a volume of special purpose pipes which collectively are treated as a contiguous region.  These pipes are much like regular pipes, except that they consume an arbitrary region of the $4\times 4\times 4$ cell.

\subsection{Braid synthesis}
As defined, the braid compaction problem takes an arbitrary braid as input.  Thus our algorithm need not address the synthesis of a quantum circuit into a braid.  Indeed, the force-directed braid model requires only that rigid collections of pipes (i.e., cuboids) be specified along with rotation and time-ordering constraints.

For concreteness, however, we will assume that the initial braid is constructed from a quantum circuit in the canonical way as described in~\secref{sec:surface-code}. That is, qubits are represented by pairs of primal defects.  Single-qubit preparation corresponds to two bent pipes connected to form a ``U'' shape and single-qubit measurement is the same, except that the U shape is upside-down.  Hadamards, state distillation and tabletop measurements are abstracted as cuboids of fixed dimension. CNOT gates are constructed by wrapping a dual loop around corresponding primal loops.
%The resulting initial braid is effectively a two-dimensional object that mirrors the structure of the input circuit.  It is the starting point for our compaction algorithm.

The Hadamard cuboid is three cells wide, four cells deep and four cells high. This cuboid is larger than is strictly necessary to enclose the Hadamard operation. Part of the Hadamard operation involves cutting a boundary around the corresponding logical qubit.  The volume given above provides enough room for the Hadamard operation to take place inside boundary, while enforcing that defects outside of the boundary are a safe distance away. Affixed to opposite faces of the cuboid are pairs of straight pipes representing the input and output logical qubit.

The specifics of the $T$-gate braid depend on the distillation protocol and on the desired gate accuracy, but otherwise follow~\figref{fig:tgate-time-optimal}.  Our compaction algorithm is flexible enough to allow any type of distillation scheme.  For simplicity, we will assume the existence of two cuboid regions for each $T$ gate, one for $\ket A$ and one for $\ket Y$.  Straight pipes representing the output are affixed to the top of each cuboid.

\subsection{Gravity}
The primary ``force'' in the algorithm is a vector field that loosely resembles physical gravity acting on the braid.  With each cell in the grid, we associate two vectors of the form $(a,m)$, specified by an axis $a \in \{x, y, z\}$ and a magnitude $m \in \mathbb Z$.  The first vector represents a force on the primal pipe contained in the cell, and the second vector represents a force on the dual pipe.

\begin{figure}
\centering
\includegraphics[scale=.2]{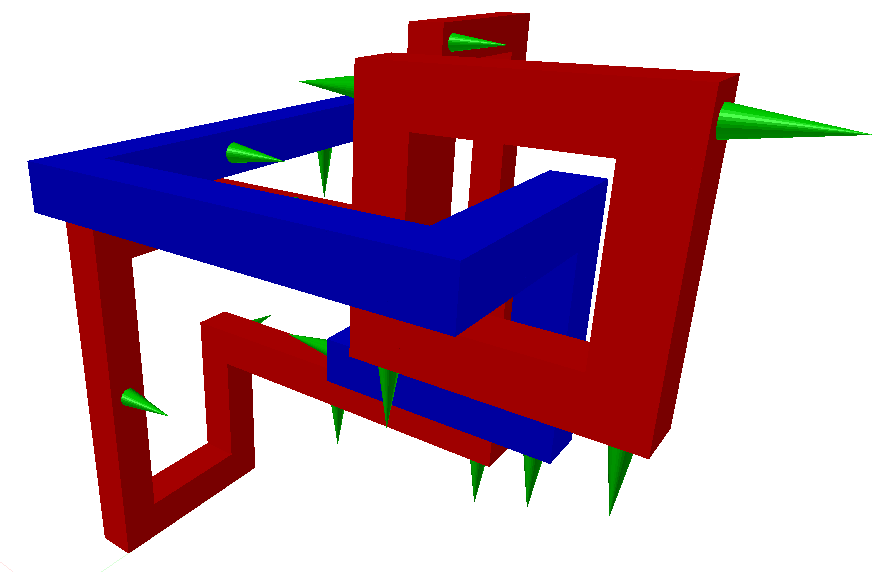}
\caption{\label{fig:gravity-example}
Gravity vectors (shown as green cones) generally point downward, but may point in any direction.}
\end{figure}

There are a number of reasonable ways to initialize and update the gravity field as defects are moved around.  The simplest strategy is to assign a fixed, negative magnitude to each spacetime point and align the vector along the $z$-axis so that the force always points downward.
In order that defects may slide past each other, though, we allow vectors to point sideways along the $x$ and $y$ axes, as well. See~\figref{fig:gravity-example}.  Roughly, gravity vectors are assigned to point to the closest cell from which the defect may then move downward.  For example, a primal pipe occupying cell $(x,y,z)$ may be blocked by a dual pipe in cell $(x,y,z-1)$.  If, however, cells $(x+1,y,z)$ and $(x+1,y,z-1)$ are empty, then the primal gravity vector for cell $(x,y,z)$ is assigned to point along the positive $x$-axis.

% Gravity magnitudes are assigned based on the height of the cell and the direction of the vector.  If the maximum height of the grid is $h$ then a gravity vector pointing along the negative $z$-axis is assigned a magnitude of $z-h+1$.  Similarly, a vector pointing along the $x$- or $y$-axis is assigned a magnitude (in absolute value) of $h-z$.  That is, the absolute value of gravity magnitude increases linearly with height.  Gravity vectors do not point in the positive $z$ direction.

\subsection{Tension}
The gravity force, while effective at directing pipes toward the bottom of the grid, has the effect of stretching strings and loops, thus increasing the length of the the braid.  This happens, for example, when a loop is pulled by gravity in one direction but a small segment of the loop is prevented from moving because other defects are in the way.
When a string or loop becomes very long, it may take up space that could otherwise be occupied by other parts of the braid.  To prevent this behavior we implement a tension force which acts to reduce the length of a string.

Tension is applied to each string of defects independently.  For each pipe in the string, there is a force pulling in the direction of the input face and a force pulling in the direction of the output face.  For example, a pipe connected to the $-x$ and $+z$ faces will experience a force in the $-x$ and $+z$ directions.  The magnitude of the force is proportional to the length of the string, just as for a physical spring.

\begin{figure}
\centering
\begin{subfigure}[b]{.45\linewidth}
  \centering
  \includegraphics[scale=.4]{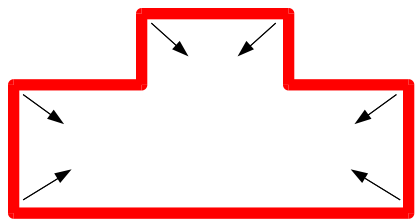}
  \caption{before}
\end{subfigure}
\begin{subfigure}[b]{.2\linewidth}
  \centering
  \includegraphics[scale=.4]{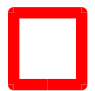}
  \caption{after}
\end{subfigure}
\caption{\label{fig:tension}
The tension force pulls inward on each of the corners of the loop. The result is a smaller rectangular loop.
}
\end{figure}

This choice of tension forces means that the inward and outward forces cancel for straight pipes.  Bent pipes, however, feel an inward force toward the rest of the string.  This inward force tends to decrease the curvature of the string, thereby reducing its length. See~\figref{fig:tension}.

Tension forces also act on cuboids. Each of the pipes connected to a cuboid exerts a force that pulls in the direction of the pipe.  Again, the force is proportional to the length of the string to which each connecting pipe belongs.

\subsection{Compaction}
The braid is initially placed above the three-dimensional grid.  Since the braid may be wider than the grid dimensions, a funnel is placed on top of the grid.  This allows the braid to slowly deform according to the geometry of the lattice.
Compaction then proceeds by iterating through each of the cuboids and strings.  Cuboids are translated or rotated as a single rigid object. Other regions of pipes form strings which either connect to cuboids or form loops. Strings are treated as flexible objects in which each pipe can be translated independently.

Associated with each pipe is a velocity vector.  Each pipe in a string is moved by first taking the initial velocity vector and updating it according to the gravity and tension forces at that location.  The pipe is then translated according to the direction and magnitude of the new velocity vector.  During the move, additional pipes may be added or removed in order to maintain connectivity of the string.
  
To translate a cuboid, the total velocity is calculated by summing each of the individual velocity vectors.  Similarly, the gravity and tension forces are calculated by summing the force vectors associated with each pipe.  The cuboid velocity is then updated by dividing the total force by the number of pipes (each pipe is assumed to have the same mass) and then adding to the existing velocity.  Finally, the cuboid is translated according to the direction and magnitude of the velocity vector.

In the case of tabletop measurement translations along the $z$-axis, we must preserve the partial ordering.  When translating a tabletop $m$ along the $-z$-axis we must check the height of the other measurements on which $m$ depends.  Likewise, when translating $m$ along the $+z$-axis, we must check the height of the measurement that depends on $m$.

Cuboid rotations are performed similarly by calculating a rotational velocity according to the moments of each pipe and the torque due to gravity and tension forces.  Rotation about a given axis is performed only if rotation is allowed and the magnitude of the angular velocity is large enough to induce a rotation of $\pi/2$.

Of course, all of the moves performed during compaction must maintain the braid topology.  In particular, we do not allow pipes to intersect nor do we allow a pipe of one type to pass through a pipe of opposite type in order to arrive at its destination. Though we do allow defects of the same type to pass through each other since this does not change the computation. It is possible for the translational or rotational path of a group of pipes to be blocked by other pipes.  When this happens, we say that a collision has occurred.

Collisions are resolved by first calculating the velocity and mass of each of the two objects involved.  In the case that a cuboid collides into multiple pipes, the impeding pipes are treated as a collective object.  The velocities of the two objects are then recalculated according to the equations of motion for a partially inelastic collision.
In this way, distinct parts of the braid are able to communicate with each other.  For example, large objects may shift smaller objects out of the way and linked loops may tug on each other. However, the rules for moving each pipe are still entirely local and relatively simple.

Note that a collision can also occur between time-dependent measurements even when the two cuboids are not located nearby each other.  Such a collision happens if the vertical motion of one of the measurements would cause a violation of relative time-ordering constraints.  The collision is non-local, but can be efficiently identified and resolved by maintaining a dependency tree with the location of measurement.

Since the topology of the braid is preserved at each step, compaction can be terminated at any time. Indeed, there are a number of reasonable termination conditions.  Compaction can be stopped after a fixed number of iterations, or a fixed amount of time.  It can also be stopped when all of the pipes are located below a particular height, or as soon as all of the pipes fit within the dimensions of the lattice.  The termination condition could also be more complicated.  For example, compaction could be halted if the maximum height remains unchanged for a certain fixed number of iterations.

\subsection{Performance and scalability}
\label{sec:performance-scalability}
The complexity of a single compaction iteration scales as the size of the braid. The size of a canonical braid is $O(nm)$ where $n$ is the number of qubits and $m$ is the number of gates in the input circuit.  The number of iterations required to obtain good compaction results depends on the ratio of the lattice area---i.e., the $x$-$y$ plane---to the braid size.  In the case that the lattice area is large compared to the braid size, it seems reasonable to expect the braid to flatten in time proportional to the height of the canonical braid.  If the canonical braid has area large compared to the lattice, then $O(nm)$ iterations may be required in order to funnel then entire braid into the proper bounding box.

For small circuit sizes, a runtime of $O(n^2m^2)$ is reasonable.  But for large circuits consisting of thousands of qubits and possibly millions or billions of gates, we require a better strategy. Indeed, we cannot hope to globally optimize braids for large-scale problem sizes.  Instead, the circuit is partitioned into subcircuits of manageable size and the braid is synthesized and compacted hierarchically.  Just as we treat single-qubit Hadamards as atomic cuboids of fixed size, we may consider sub-braids as fixed size cuboids.

Each sub-braid is represented as a tangle of defects in which some defects are anchored to grid boundaries.  Subject to the anchoring constraints, the sub-braid is compacted as normal.  Once its compacted size is determined, the sub-braid is then treated as a black-box in the larger braid.
If two sub-braids contain measurements that are time-ordered, then the sub-braids must also be time ordered.  But again, this is no different than time ordering restrictions on tabletop measurements in the original model.

We anticipate that the best partitioning strategy will be one that reflects the structure of the input circuit.  Reasonable representations of large input circuits will be hierarchical and it should be possible to mimic this hierarchy for large-scale compaction.
This technique will be particularly useful for highly repetitive circuits.  Repeated sub-circuits can be synthesized and compacted once, and then duplicated in the larger braid. 

\subsection{Implementation and results}
We have implemented the force-directed compaction algorithm in C++ as a tool called Braidpack.  Braidpack takes, as input, a representation of a circuit along with physical space restrictions.  It produces, as output, a compact logically equivalent surface code braid.  

\begin{figure}
\centering
\hfill
\begin{subfigure}[b]{.49\linewidth}
  \centering
  \includegraphics[scale=.25]{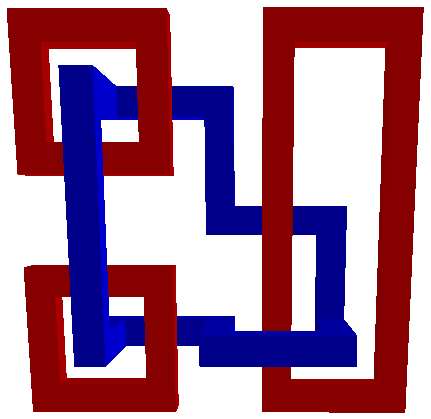}
  \caption{}
\end{subfigure}
\hfill
\begin{subfigure}[b]{.49\linewidth}
  \centering
  \includegraphics[scale=.25]{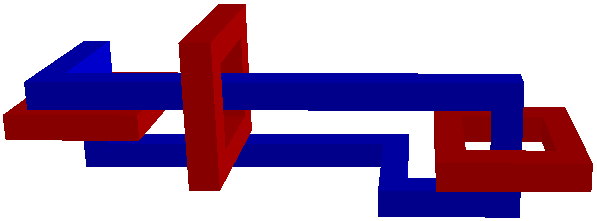}
  \caption{}
\end{subfigure}
\caption{\label{fig:cnot-compaction}
Compaction of a single CNOT using a prototype of the force-directed algorithm. (a) A canonical CNOT braid is initially arranged vertically. (b) After compaction, the braid has been almost completely flattened.}
\end{figure}

The current implementation is not yet fully functional, but is capable of synthesizing and compacting arbitrary circuits of CNOT gates, including qubit preparation and measurement. \figref{fig:cnot-compaction} shows the result of compaction on a single CNOT gate.  The tension force first contracts the primal loop on the right-hand-side.  Then gravity flattens the braid.  Compaction in this example was done without implementing collisions between pipes.  As a result, tension is unable to fully contract the dual loop.  Once implementation is complete, we expect the braid to fully flatten and contract.  

\begin{figure}
\centering
% \begin{subfigure}[b]{.3\linewidth}
%   \includegraphics[width=\linewidth,angle=90]{images/y-distillation-circuit}
%   \caption{}
% \end{subfigure}
\hfill
\begin{subfigure}[b]{.45\linewidth}
  \centering
  \includegraphics[width=.6\linewidth]{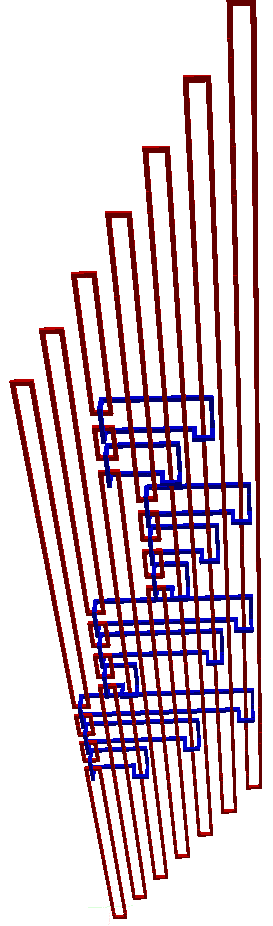}
  \caption{}
\end{subfigure}
\hfill
\begin{subfigure}[b]{.45\linewidth}
  \centering
  \includegraphics[width=.6\linewidth]{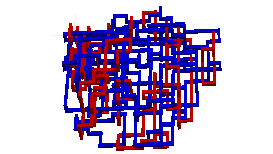}
  \caption{}
\end{subfigure}
\caption{\label{fig:y-distillation-compaction}
Compaction of eleven CNOT gates with a prototype implementation of the force-directed algorithm.
The canonical braid (a) is compressed into a smaller but topologically equivalent braid (b).}
\end{figure}

\figref{fig:y-distillation-compaction} shows the same prototype implementation of Braidpack for a circuit composed of eleven CNOT gates.  For simplicity of implementation, the qubit preparation and measurements in the canonical braid are arranged in a staircase fashion.  Ignoring the staircases, the canonical braid has a bounding box of size ($3 \times 16 \times 34$), whereas the the compacted braid fits in a bounding box of size ($10 \times 13 \times 6$), a factor of four improvement along the time axis.  Again, we expect improved results once the Braidpack completed.

In order to facilitate debugging, we have developed a braid visualization tool called Braidview.  This tool creates a single file from a braid or sequence of braids.  The file can be viewed in Blender~\cite{Blender}, a third-party open-source $3$-D modeling application.  Braidview is capable of separately rendering primal and dual defects (as in~\figref{fig:y-distillation-compaction}), as well as gravity vectors (see~\figref{fig:gravity-example}). The backbone of Braidview is a set of rendering functions that use the Blender Python-API.  These, and some additional functions, are used by similar visualization tools Nestcheck and Autotune~\cite{Milburn2012,Fowler2012a}.

\section{Compaction by simulated annealing}
\label{sec:simulated-annealing}
In this section we describe our second compaction algorithm, which is based on simulated annealing. Simulated annealing is a general optimization technique that has been applied to a wide variety of problems.  The main idea is to explore the solution space by hopping randomly from the current solution to a nearby solution. Hops that result in an improved solution are kept.  In order to avoid local minima, hops that result in a less desirable solution are also kept with some non-zero probability, thus permitting broader exploration of the set of possible solutions.

Our simulated annealing algorithm is based largely on a procedure used for VLSI placement \cite{Hsieh}.  In the VLSI algorithm, circuit elements and wires are represented by rectangles.  Size, distance and connectivity constraints are given by linear inequalities on the coordinates of each rectangle.  Rectangles can be shifted around by swapping linear constraints.  The idea for braids is similar.  Defects are represented by cuboids. Size, distance and topology constraints are given by linear inequalities which can be swapped to perform topological deformation.

\subsection{Definition of the braid}
In the force-directed algorithm, the braid was modeled as a connected configuration of plumbing pieces.  Some collections of pipes formed rigid cuboids.  Other collections of pipes formed flexible strings and loops.
For simulated annealing, we take a different approach.  Each cuboid is represented by a pair of points $(p, p')$ in the three-dimensional lattice.  Point $p$ specifies the point closest to the origin (lower-left corner) and $p'$ specifies the point furthest from the origin (upper-right corner).
Defect strings and loops are also represented by cuboids.  A string of defects is given by a set of overlapping cuboids of arbitrary dimension.  By connecting cuboids it is possible to construct any desired loop or string.

Thus the entire braid is specified by a set of cuboids.
A \emph{layout} of $n$ cuboids is defined by $2n$ three-dimensional integer coordinates.  The $x$, $y$, $z$ dimensions of the layout are defined by the maximum $x$, $y$, and $z$ coordinates respectively.
The layout must satisfy a set of constraints which we group the constraints into the following types:
\begin{enumerate}
  \item size constraints,
  \item time-ordering constraints,
  \item minimum distance constraints,
  \item jog constraints,
  \item connectivity constraints and
  \item topological constraints.
\end{enumerate}
Except for the topological constraints, all of the constraints can be directly expressed as sets of linear inequalities.

\subsubsection{Size constraints}
\label{sec:size-contraints}
Minimum dimension constraints of a cuboid are specified by a triple $(\delta_x, \delta_y, \delta_z)$ of non-negative real values and three linear inequalities:
\begin{equation}\begin{split}
\label{eq:cuboid-dimension-constraints}
  x + \delta_x &\leq x' \\
  y + \delta_y &\leq y' \\
  z + \delta_z &\leq z'
  \enspace.
\end{split}
\end{equation}
For string cuboids (those that are not $H$ gates or table-like measurements), $\delta_x = \delta_y = \delta_z = d/4$, where $d$ is the code distance. Hadamard and $T$ gates may be rotated $90$ degrees about the $z$-axis. 
Each gate can take on one of four different rotations $\{0, \pi/2, \pi, -\pi/2\}$.
Rotations $0$ and $\pi$ correspond to the set of constraints given by~\eqnref{eq:cuboid-dimension-constraints}.  
Rotations $\pm \pi/2$ correspond to the same set of constraints in which $\delta_x$ and $\delta_y$ have been exchanged.

We therefore assign one of two sets of constraints to each $H$ and $T$ gate, either the constraints of~\eqnref{eq:cuboid-dimension-constraints} or the permuted version. The corresponding cuboids must satisfy all constraints from at least one of sets.

\subsubsection{Time-ordering constraints}
The non-deterministic implementation of $T$ gates in the surface code induces a partial time-ordering of tabletop measurement regions. As discussed in~\secref{sec:surface-code}, this partial ordering requires that, for certain pairs, one tabletop measurement must be located above another tabletop measurement.
The time-ordering constraint for two dependent measurements, $a$, $b$ is given by,
\begin{equation}
  z'_a + 1 \leq z_b
  \enspace .
\label{eq:time-ordering-constraint}
\end{equation}

\subsubsection{Minimum distance constraints}
\label{sec:min-distance-constraints}
Like the size constraints, minimum distances are proportional to $d$, the distance of the code. With a few exceptions (see \secref{sec:jog-nodes} and \secref{sec:connectivity-constraints}), primal defect cuboids must be at least a distance $d$ away from other primal defects. Similarly, dual defect cuboids must be $d$ away from other dual cuboids.  Cuboids of opposite type must be at least $d/4$ apart.

If two cuboids $r_i, r_j$ must be separated by $\delta$, then at least one of the following constraints must be satisfied:
% \begin{equation}
% \begin{tabular}{ccc}
%   $x'_i - x_j \geq \delta$ & $x_i - x'_j + y_i - y'_j \geq \delta$ & $x_j - x'_i + y_j - y'_i + z_j - z'_i \geq \delta$  \\
%   $x'_j - x_i \geq \delta$ & $x_j - x'_i + y'_i - y'_j \geq \delta$ & $x_i - x'_j + y_j - y'_i + z_j - z'_i \geq \delta$ \\
%   $y'_i - y_j \geq \delta$ & $x_j - x'_i + y_j - y'_i \geq \delta$ & $x_i - x'_j + y_i - y'_j + z_j - z'_i \geq \delta$ \\
%   $y'_j - y_i \geq \delta$ & $x_i - x'_j + y_j - y'_i \geq \delta$ & $x_j - x'_i + y_i - y'_j + z_j - z'_i \geq \delta$ \\
%   $z'_i -z_j \geq \delta$ & $x_i - x'_j + z_i - z'_j \geq \delta$ & $x_j - x'_i + y_j - y'_i + z_i - z'_j \geq \delta$ \\
%   $z'_j - z_i \geq \delta$ & $x_j - x'_i + z_i - z'_j \geq \delta$ & $x_i - x'_j + y_j - y'_i + z_i - z'_j \geq \delta$ \\
%   & $x_j - x'_i + z_j - z'_i \geq \delta$ & $x_i - x'_j + y_i - y'_j + z_i - z'_j \geq \delta$\\
%   & $x_i - x'_j + z_j - z'_i \geq \delta$ & $x_j - x'_i + y_i - y'_j + z_i - z'_j \geq \delta$ \\
%   & $y_i - y'_j + z_i - z'_j \geq \delta$ & \\
%   & $y_j - y'_i + z_i - z'_j \geq \delta$ & \\
%   & $y_j - y'_i + z_j - z'_i \geq \delta$ & \\
%   & $y_i - y'_j + z_j - z'_i \geq \delta$ &
% \end{tabular}
% \end{equation}
\begin{equation}
\begin{tabular}{ll}
  $x'_i +\delta \leq x_j$ & $x'_j + \delta \leq x_i$ \\
  $y'_i +\delta \leq y_j$ & $y'_j + \delta \leq y_i$ \\
  $z'_i +\delta \leq z_j$ & $z'_j + \delta \leq z_i$ \\
\end{tabular}
\label{eq:min-distance-constraints}
\end{equation}
Each constraint corresponds to a different relative arrangement of the two cuboids.  The $x'_i +\delta \leq x_j$ constraint, for example, enforces that $r_i$ is placed to the left of $r_j$. Whereas $z'_i +\delta \leq z_j$ requires that $r_i$ be placed below $r_j$.

\subsubsection{Jog nodes}
\label{sec:jog-nodes}
A fixed string of defects may be represented by a set of overlapping cuboids each of which has a fixed orientation along one of the three axes.  However, in order to accommodate topological deformation we require a representation that allows for flexible strings of cuboids. This is analogous to a VLSI instance in which an arbitrary number of jogs are allowed in each wire.  To fulfill this requirement, we introduce an object called a \emph{jog node}.

A jog node is a set of six cuboids, each of which has a particular orientation axis.  The first cuboid is oriented along the $+x$ axis, the second along the $+y$ axis, and the third along the $+z$ axis.  The fourth, fifth and sixth cuboids are oriented along the $-x$, $-y$ and $-z$ axes, respectively.  Each cuboid in the jog node is allowed to expand along its corresponding axis.  Adjacent cuboids are required to overlap so that the entire jog node forms a continuous path.  The constraints for a jog node are given by:
\begin{equation}
\begin{tabular}{cccccc}
  $x_1 \leq x_2$, &$y_1 = y_2$, &$z_1 = z_2$, &$x_1' = x_2'$, &$y_1' \leq y_2'$, &$z_1' = z_2'$,\\
  $x_2 = x_3$, &$y_2 \leq y_3$, &$z_2 = z_3$, &$x_2' = x_3'$, &$y_2' = y_3'$, &$z_2' \leq z_3'$,\\
  $x_3 \geq x_4$, &$y_3 = y_4$, &$z_3 \geq z_4$, &$x_3' = x_4'$, &$y_3' = y_4'$, &$z_3' = z_4'$,\\
  $x_4 = x_5$, &$y_4 \geq y_5$, &$z_4 = z_5$, &$x_4' \geq x_5'$, &$y_4' = y_5'$, &$z_4' = z_5'$,\\
  $x_5 = x_6$, &$y_5 = y_6$, &$z_5 \geq z_6$, &$x_5'=x_6'$, &$y_5' \geq y_6'$, &$z_5' = z_6'$.
\end{tabular}
\label{eq:jog-node-constraints}
\end{equation}

It possible to connect two jog nodes at their endpoints.  Given the sixth cuboid $a6$ of jog node $a$ and the first cuboid $b1$ of jog node $b$ the endpoints are connected by requiring
\begin{equation}
  x_{a6} = x_{b1}, y_{a6} = y_{b1}, z_{a6} = z_{b1}
  \enspace .
\end{equation}
In this way, jog nodes can be connected to form an arbitrary defect path of any length.  It is possible to form both loops and open ended strings.

\def\imagetop#1{\vtop{\null\hbox{#1}}}
\begin{figure}
 \centering
\begin{tabular}{cccc}
  &
  \includegraphics[scale=.25]{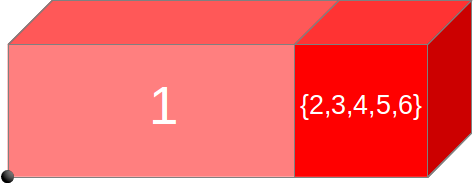} &
  \includegraphics[scale=.25]{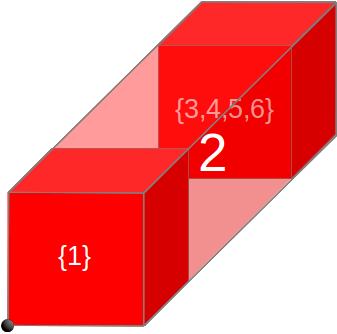} &
  \includegraphics[scale=.25]{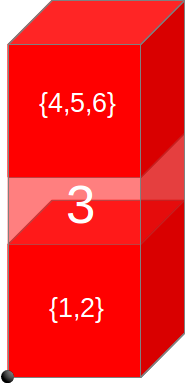}\\
  \raisebox{-2.5cm}{\includegraphics[scale=.5]{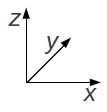}} &
  \imagetop{\includegraphics[scale=.25]{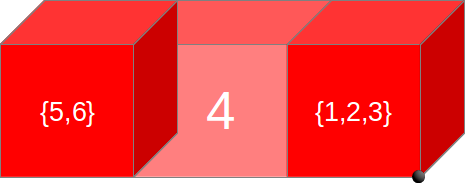}} &
  \imagetop{\includegraphics[scale=.25]{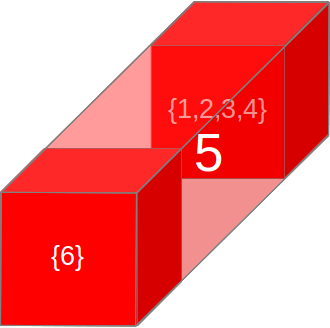}} &
  \imagetop{\includegraphics[scale=.25]{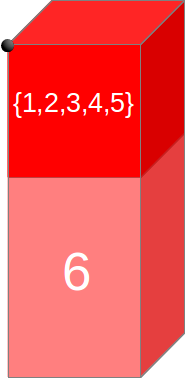}}
\end{tabular}
 \caption{\label{fig:jog-node-loop}
   A jog node consists of six overlapping cuboids.  Each cuboid is allowed to extend in only one direction, and only one cuboid in the node may be extended.  The six possible jog node configurations are shown above.  The node origin is indicated by a black dot, where visible.
}
\end{figure}

The jog node constraints, as stated, conflict with the minimum distance constraints in~\secref{sec:min-distance-constraints}.  For example, cuboids $a1$ and $a2$ are required by~\eqnref{eq:jog-node-constraints} to be connected, but are required by~\eqnref{eq:min-distance-constraints} to be separated by $\delta$. As a workaround, we first require that each jog node be oriented along at most one axis.
This is accomplished by changing the appropriate inequality constraints to equality constraints.  For example, to force an orientation along the $+x$ axis only, leave the $x_1 \leq x_2$ constraint alone and change all of the other inequalities to equalities.
Then the cuboid corresponding to the $+x$ axis can be of arbitrary size (subject to minimum dimension constraints) and all other cuboids of the node must fit inside of it.
See~\figref{fig:jog-node-loop}.

Next, remove the minimum distance constraints for all jog node cuboids except those that correspond to the orientation axis.  Finally, remove minimum distance constraints between cuboids in adjacent jog nodes. Now, overlapping cuboids within the same jog node or between connected jog nodes are consistent with all other constraints.

A jog node may also be configured to take no orientation.  In this case, all cuboids in the node are constrained to be of minimum size, i.e., $x + \delta_x = x'$, $y + \delta_y = y'$, $z + \delta_z = z'$.  Furthermore, all minimum distance constraints involving the node are removed.  This type of node will either be unconnected to any other node (in which case it can be removed), or it will be contained entirely within another jog node.  In either case, its distance from other objects in the braid is unimportant.

\subsubsection{Connectivity constraints}
\label{sec:connectivity-constraints}
Jog nodes allow for arbitrary defect paths and loops.  We must also define how jog nodes are used to connect to cuboids such as Hadamard gates and state distillation.
Each gate cuboid contains some number of ports to which string defect cuboids are allowed to attach. The locations of the ports are fixed relative to the gate.  However, since gates can be rotated, the constraints that describe the connection must correspond to the permutation of the dimensional constraints from~\secref{sec:size-contraints}.

\begin{figure}
\centering
\includegraphics[height=4.5cm]{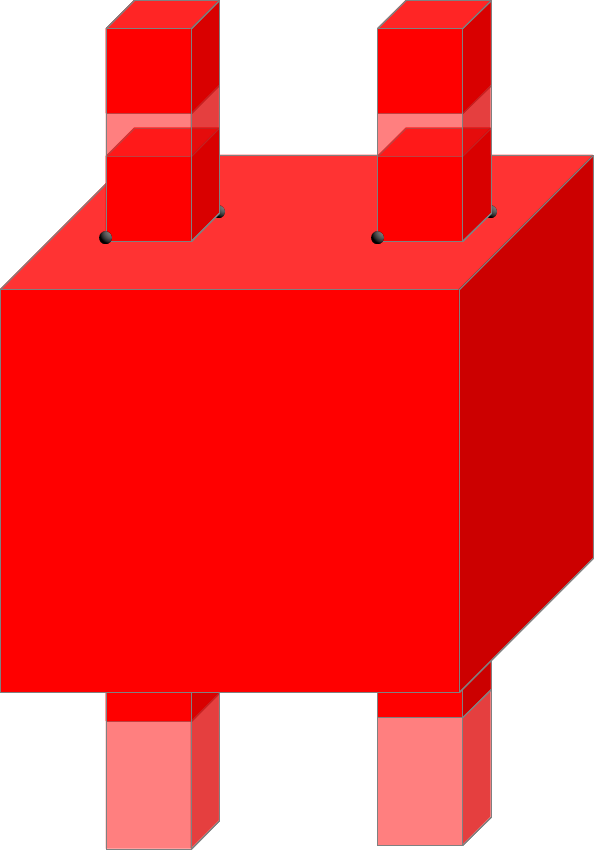}
\caption{\label{fig:gate-with-ports}
The above cuboid has four ports defined on its surface, two on top and two on the bottom.  Jog nodes are affixed to the points that define each port.
}
\end{figure}

A port is a rectangle defined by two coordinates on the surface of the gate.  A jog node is connected to a port by requiring that certain coordinates of the jog node cuboid match the coordinates of the port. For example, if the input port $(x, y, z)$, $(x', y', z)$ is located on the top of the gate, then the jog node connection constraint is given by
\begin{equation}
  x_3 = x, y_3 = y, x'_3 = x', y'_3 = y', z_3 = z
  \enspace .
\end{equation}
See~\figref{fig:gate-with-ports}.

To maintain consistency, the minimum distance constraints between the gate and the connecting jog node must be eliminated.  Note that it is still possible for two connected gates to achieve a separation of exactly $d$.  In this case, the node connected to the output port of the first gate is also connected to the input port of the second gate, and vice versa.  But since each node is of minimum size, the minimum distance constraints between the node and the gates do not apply (see~\secref{sec:jog-nodes}).

\subsubsection{Topological constraints}
Finally we address the topological constraints.  Informally, these constraints enforce the linking between loops.  Links between loops of the same type are trivial and need not be constrained.  However, certain linking properties between loops of different types must be maintained. In particular, it is sufficient to consider the linking number for each primal-dual loop pair. For each primal-dual pair $(l_p, l_d)$ we have the following constraint
\begin{equation}
 l_{pd} = L_{pd} \mod 2
\end{equation}
where $l_{pd}$ is the linking number of loops $l_p$ and $l_d$ and $L_{pd} \in \{0,1\}$ is an input parameter.

There is a simple linear-time algorithm to compute the linking number between two loops (see, e.g., \cite{Kauffman2001}).  However, in order to efficiently compute the cost function of a layout, we will require that all constraints be linear. See~\secref{sec:annealing-algorithm}.

We impose linear topology constraints separately for loop pairs with odd linking number (i.e., loops that are linked) and loop pairs with even linking number (loops that are not linked).  First consider two loops with odd linking number.  One of the loops consists of primal defects and the other loop consists of dual defects.  To the primal loop, attach a new primal cuboid which we will call a \emph{linking node}.  The linking node has dimension $(5d/4\times d/4\times 5d/4)$. It is attached to the primal loop by connecting one of the jog nodes to the top and connecting an adjacent jog node to the bottom.

The linking node is also attached to jog nodes of the dual loop.  Instead of connecting on the top, the dual jog nodes are connected on either side of the linking node.  The dimensions of the linking node are about twice as large as would otherwise be necessary for maintaining minimum distance constraints between the primal and dual cuboids.  The extra space is used as a placeholder.

As the simulated annealing algorithm proceeds, the linking number between the two loops may change.  The jog nodes that were originally connected to the linking node must remain connected.  But other cuboids from the loops are unrestricted and may cross each other.  At the end of the algorithm the linking node is removed leaving some empty space.  

\begin{figure}
\centering
\begin{subfigure}[b]{.25\linewidth}
  \includegraphics[width=.7\linewidth]{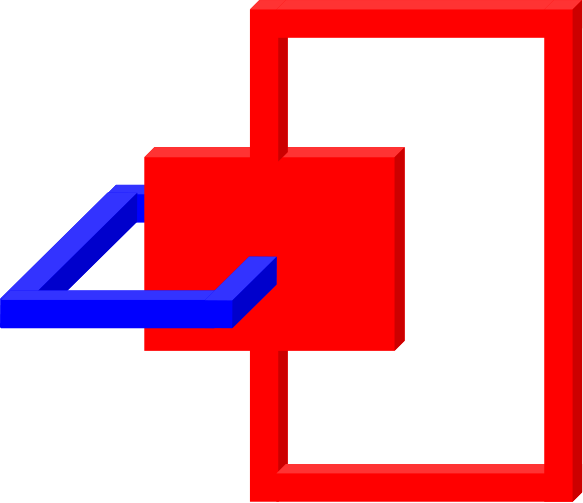}
  \caption{}
\end{subfigure}
\begin{subfigure}[b]{.25\linewidth}
  \includegraphics[width=.7\linewidth]{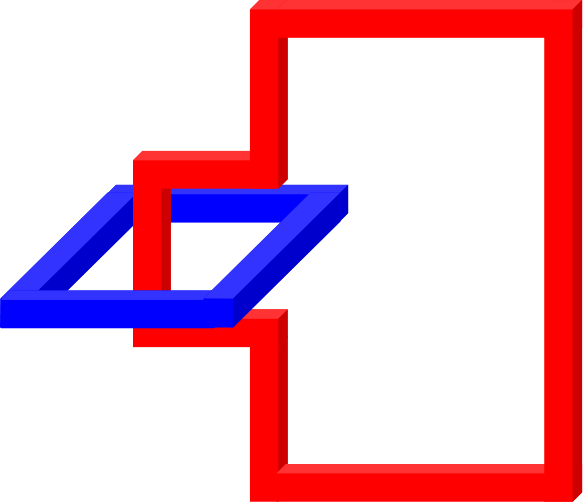}
  \caption{}
\end{subfigure}
\begin{subfigure}[b]{.25\linewidth}
  \includegraphics[width=.7\linewidth]{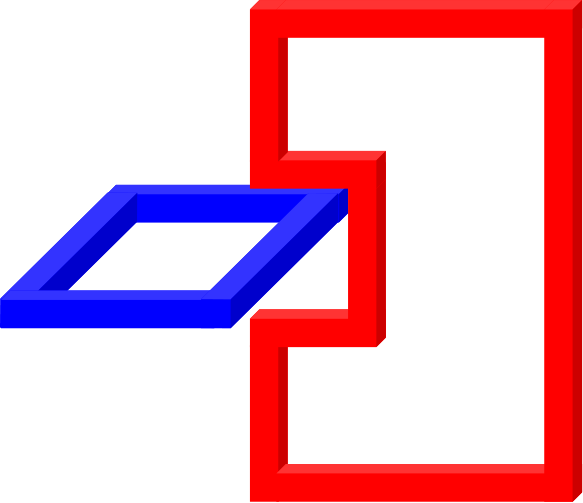}
  \caption{}
\end{subfigure}
\caption{\label{fig:linking-node}
When a primal and a dual loop are linked in the canonical braid a linking node (a) is inserted and attached to both loops.  Once compaction has completed, the linking node is removed.  The linking number can be left unchanged (b), or toggled (c) if necessary.}
\end{figure}

The primal and dual loops must now be reconnected.  However, we have a choice.  We may either connect the dual loop so that it is inside of the primal loop.  Or we may connect the dual loop so that it is outside of the primal loop.  In effect, the choice of reconnection determines whether the linking number is even or odd.  We may simply choose the configuration that yields an odd linking number.
See~\figref{fig:linking-node}.

Now consider a primal loop and a dual loop with even linking number.  Since these loops are unlinked, they may be far apart in spacetime.  Thus the linking node strategy is not practical.  However, we must still ensure that these loops remain unlinked in the output of the algorithm.  We will do this by requiring that the dual loop remain sufficiently far from the primal loop at all times.

Consider the primal loop.  It is composed of a set of connected primal cuboids, some of which are jog nodes and some of which are Hadamard or state distillation cuboids.  Let $x$ be the minimal $x$-coordinate of any cuboid in this set and let $x'$ be the maximal $x$-coordinate of any cuboid in the set.  Similarly define $y$, $y'$, $z$ and $z'$ as the minimal and maximal $y$- and $z$-coordinates.  Then the entire primal loop is contained in a bounding box of dimension $(x'-x, y'-y, z'-z)$.

\begin{figure}
\centering
\includegraphics[scale=.3]{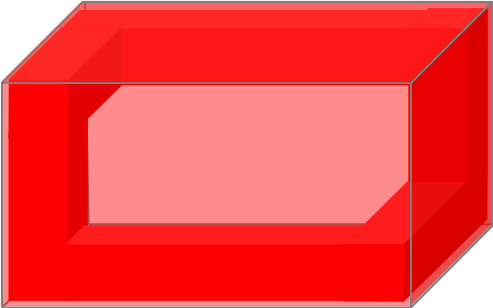}
\caption{\label{fig:enclosing-cuboid}
In order to avoid unwanted links, a cuboid is placed around each primal loop.  Dual loops which do not link with the primal loop are prohibited from entering the enclosing cuboid.
}
\end{figure}

If all of the cuboids in the dual loop stay outside of the bounding box that encloses the primal loop, then the linking number is guaranteed to be zero. We therefore introduce a new cuboid that encloses the primal loop. For all dual loops which have even linking number with the corresponding primal loop, we add primal-dual minimum distance constraints between the dual cuboids and the enclosing cuboid. See~\figref{fig:enclosing-cuboid}.

In order to ensure that the new cuboid actually encloses the primal loop, additional variables and constraints are required.  Let $(x,y,z)$ and $(x',y',z')$ be variables describing the enclosing cuboid.  Then for each cuboid $(x_i, y_i, z_i)$, $(x'_i, y'_i, z'_i)$ in the primal loop we require that
\begin{equation}
\begin{tabular}{ll}
  $x \leq x_i$ & $x'_i \leq x'$ \\
  $y \leq y_i$ & $y'_i \leq y'$ \\
  $z \leq z_i$ & $z'_i \leq z'$.
\end{tabular}
\end{equation}

\subsection{The annealing algorithm}
\label{sec:annealing-algorithm}
The algorithm takes a canonical braid as input.  Initialization consists of constructing all of the cuboids and constraint groups. The instance includes a set of coordinates $P$, which can be divided into sets of integers $X$, $Y$ and $Z$ corresponding to the $x$-, $y-$, and $z$-coordinates, respectively.  The constraints can be represented as a set $C$ of integer triples. Some of the constraints, such at time-ordering constraints, must be satisfied for all possible layouts.  Other constraints may be partitioned into subsets for which the layout must satisfy at least one of the constraints in the subset.
Let $C'$ be the set of all constraints that must always be satisfied, let $C''$ be the remaining constraints and let $B$ be the corresponding partition into constraint subsets.  Let $A \subset C$ be the set of ``active'' constraints such that $C' \subset A$ and $A$ contains exactly one constraint from each element of $B$. 

Start by choosing a set of active constraints such that all constraints in $A$ are satisfied by the canonical braid.
The algorithm then proceeds by repeating the following sequence.
\begin{enumerate}
  \item Randomly select an element $\beta \in B$.
  \item Randomly select a constraint $b \in \beta$ such that $b \not\in A$.
  \item Locate the single constraint in $b' \in A \cap \beta$.  Remove $b'$ from $A$ and replace it with the new constraint $b$.
  \item Compute the new minimum bounding box size and corresponding cost function.
  \item If the new set of active constraints is infeasible, then reject the swap by removing $b$ from $A$ and replacing with $b'$.
  \item If the cost is smaller than before, keep the new constraint.
  \item If the cost is larger than before, then keep the new constraint with probability given by the annealing schedule (see below).
\end{enumerate}

In order for the algorithm to be efficient, we require an efficient way to compute the size of the minimum bounding box.  This can be done using the constraint graph method proposed in~\cite{Liao1983} and used by~\cite{Hsieh}.  First, partition the active constraints into three sets: those that involve only $x$ coordinates, those that involve only $y$ coordinates and those that involve only $z$ coordinates.  Note that there are no constraints that involve coordinates for two different axes.  Consider just the set of $x$-coordinates $X$. We construct a weighted directed graph $G_X = (V_X, E_X)$.  Assign $V_X = X \cup \{x_\emptyset, x_\infty\}$ where $x_\emptyset$ and $x_\infty$ are a boundary coordinates. For each constraint $x_i \leq x_j + d_{ij}$ there is a directed edge from vertex $x_i$ to vertex $x_j$ with weight $d_{ij}$.  The value of each coordinate $x \in X$ is assigned by computing the longest path from $x_\emptyset$ to $x$.  Assuming that the set of constraints can be satisfied, $G_X$ is a acyclic.  Thus the longest path can be computed in linear time by negating the weights and using Dijkstra's algorithm.  Constraint graphs for $y$ and $z$ coordinates are similarly constructed.

The cost of constructing the initial constraint graphs is $O(n^2)$, where $n$ is the number of cuboids.  Once the graphs are constructed, updates can be computed by an online algorithm.  When a constraint swap is performed, only those paths affected by the corresponding vertices need to be recalculated. This algorithm can also detect cycles induced by the new constraint.  If a cycle is detected, then the set of constraints is infeasible and the swap is rejected.

There are a number of choices of cost function.  The goal is to construct a braid of small height that fits into an $x$-$y$ area of fixed size.  The first step is to ensure that the braid fits into that area.  We initially set the cost function as the $x$-coordinate of the bounding box.  Once this $x$-coordinate is small enough, we impose a global constraint that the $x$-coordinates of all cuboids must be no greater than that of the bounding box. We then set the cost function as the $y$-coordinate of the bounding box and repeat the procedure.
Finally, once the entire braid fits into the $x$-$y$ area, we minimize over the height.

For VLSI placement Hsieh, Leong and Liu use a fixed-ratio temperature schedule in which the temperature is reduced by a constant factor after each time step.  This schedule is simple and efficient and can also be used for our algorithm.  Other kinds of schedules could also be used.

\section{Discussion and future work}
The surface code provides a unique opportunity for fault-tolerant quantum circuit optimization by topological deformation.  We have defined the problem of braid compaction subject to geometric constraints, and given two heuristic algorithms.  Our tool Braidpack implements the first of these, the force-directed algorithm. Small examples indicate that compaction algorithms can lead to significant improvement in spacetime overhead when compared to the canonical braid.

Braidpack is a work in progress, and we intend implement all of the features outlined in~\secref{sec:iterative-forcing} and to scale up the size of the input examples.
We would also like to implement the simulated annealing algorithm in order to compare the performance of the two algorithms.  Indeed, we could also construct a hybrid algorithm which incorporates both techniques.

Our simulated annealing algorithm is inspired from a similar algorithm for VLSI placement.  VLSI also offers a number of other techniques including, genetic algorithms, numerical and partitioning algorithms, and force-directed algorithms that are distinct from our own~(see, e.g., \cite{Shahookar1991}).  
Perhaps some of these additional techinques could be adapted to braid compaction.

Due to similarity with VLSI compaction and other packing problems, we conjecture that braid compaction is NP-complete. A formal reduction has proven elusive, however.  Thus an obvious open problem is to confirm or refute that conjecture. 

Finally, we have focused on topological deformation.  However, other non-topological braid identities exist~\cite{Fowler2012f,Raussendorf2007a}.  Optimization involving these identities has been previously done by hand, but it may be possible to incorporate non-topological techniques into an automated tool such as ours.

\section*{Acknowledgements}
AEP would like to thank Lucy Zhang and Vinayak Pathak for insightful discussions regarding complexity of braid compaction.  Thanks also to Andrew Kennings for consultation on VLSI placement algorithms.

Supported by the Intelligence Advanced Research Projects Activity
(IARPA) via Department of Interior National Business Center contract number D11PC20166. The U.S. Government is authorized to reproduce and distribute reprints
for Governmental purposes notwithstanding any copyright annotation thereon. Disclaimer: The views and conclusions contained herein are those of the authors and
should not be interpreted as necessarily representing the official policies or endorsements, either expressed or implied, of IARPA, DoI/NBC, or the U.S. Government.

\bibliographystyle{alpha-eprint}
\bibliography{library}

\end{document}